\documentclass[doublecol,figures]{epl2}

\title{
Sequence of phase transitions induced in an array of Josephson
junctions by their crossover to $\pi$-state}
\shorttitle{Phase transitions induced in a junction array by $0-\pi$
crossover}

\author{S. E. Korshunov}
\shortauthor{S. E. Korshunov}

\institute{L. D. Landau Institute for Theoretical Physics - 
Chernogolovka 142432, Russia}

\abstract{We show that the transition of Josephson junctions between the
conventional and $\pi$ states caused by the decrease in temperature
induces in a regular two-dimensional array of such junctions not just a
single phase transition between two phases with different ordering but a
sequence of two, three or four phase transitions. The corresponding phase
diagrams are constructed for the cases of bipartite (square or honeycomb)
and triangular lattices.}

\usepackage{graphicx}
\usepackage{latexsym}
\usepackage{amsmath}
\begin{document}

\maketitle

\section{Introduction}
For several decades arrays of weakly coupled superconducting islands have
been the subject of active experimental investigations \cite{rev} for many
reasons, in particular as a simple model system which allows one to study
the interplay between fluctuations, frustration, disorder and other
factors in a more controlled situation than in bulk superconductors.
However, these studies have been restricted to arrays of conventional
junctions whose energy is minimal when the phases of two superconductors
are equal to each other.

The first experimental realization of an old theoretical idea
\cite{BKS,GKI} about fabrication of so-called $\pi$-junction whose energy
is minimal when the phase difference on the junction is equal to $\pi$ was
achieved only during last decade by  Ryazanov \etal \cite{Ryaz00} who
studied superconductor-ferromagnet-superconductor (SFS) Josephson
junctions and observed a transition from the conventional state to the
$\pi$-state taking place with the decrease in temperature \cite{GKI}. The
experimental investigation of small arrays of SFS junctions started almost
simultaneously \cite{Ryaz-arr}, but insofar has been restricted to very
modest sizes \cite{Ryaz-arr-2}.

Since the fabrication of more sizable arrays of SFS junctions is
definitely a matter of the nearest future, the present letter addresses
the question what happens with a superconducting array of Josephson
junctions when the decrease in temperature induces a crossover of the
junctions to the $\pi$-state. Although one could expect (from the evident
change of the ground state structure) that this induces a single
first-order transition between two phases with different ordering, our
analysis reveals that this is never the case and in reality an array
experiences in the crossover region a sequence of two, three or even four
phase transitions each of which is related with partial or complete
destruction (or restoration) of ordering. The structures of phase diagrams
and the natures of these transitions are established both for bipartite
lattices (square and honeycomb) and for a triangular one.

\section{ Model}
An array of identical SFS junctions can be described by the Hamiltonian
\begin{equation}                                              \label{H}
    H=\sum_{({\bf jj'})}V(\varphi_{\bf j}-\varphi_{\bf j'})\,,
\end{equation}
where $\varphi_{\bf j}$ is the phase of the superconducting order
parameter on ${\bf j}$th superconducting island, the summation is
performed over all pairs of neighboring islands connected by a junction
and $V(\theta)$ is a periodic even function of $\theta$ which can have
minima both at $\theta=0$ and $\theta=\pi$. When the contacts forming a
junction have low transparency, one can keep in the Fourier
expansion of
$$
    V(\theta)=-\sum_{p=1}^{\infty}J_p\cos(p\,\theta)$$
    only the first term 
because a typical value
of $J_p$ is strongly suppressed with the increase of $p$ \cite{GKI}.

However, in a SFS junction of an appropriate width the decrease in
temperature $T$ may force the value of $J_1$ to pass through zero and
change sign \cite{CBNB}. This leads to the transition of the junction from
the conventional state [in which the deepest minimum of $V(\theta)$ is at
$\theta=0$] to the $\pi$-state (in
which the deepest minimum is at
$\theta=\pi$). Naturally, in the vicinity of $T_0$, the temperature at
which $J_1(T)=0$, one has to keep also the next term in the Fourier
expansion of $V(\theta)$,
\begin{equation}                                               \label{V}
    V(\theta)=-J_1\cos\theta-J_2\cos(2\theta)\,.
\end{equation}
In the simplest situation the decrease of $T$ leads to the change of
$J_1(T)$ from positive to negative, while $J_2(T)$ remains positive.
Our aim consists in analyzing what phase transition (or what sequence of
phase transitions) takes place in a regular array of identical SFS
junctions 
when they experience such a transition to the $\pi$-state (also known as
$0-\pi$ crossover).

\section{Bipartite lattice}
First one has to understand what would take place with the change of the
sign of $J_1$ in the absence of thermal fluctuations. For $J_{1,2}>0$ the
minimum of the Hamiltonian (\ref{H}) with interaction (\ref{V})
on any lattice is achieved when all variables $\varphi_{\bf j}$
(defined modulo $2\pi$) are equal to each other, $\varphi_{\bf j}=\Phi$.
Therefore, the ground state is characterized by $U(1)$ degeneracy related
to the simultaneous rotation of all phases.

The form of the ground state at $J_1<0$ depends on the structure of the
lattice. We start by considering the case of a {bipartite} lattice
(square or honeycomb)
and after that will discuss the more complex case of a triangular lattice.
For any bipartite lattice the problem with $J_1<0$ can be mapped onto the
problem with $J_1>0$ just by rotating half of the variables $\varphi_{\bf
j}$ by $\pi$. In particular, this immediately defines the form of the
ground state at $J_1<0$, which has the same $U(1)$ degeneracy as at $J_1>0$
but a different (two-sublattice) structure.

When $J_1=0$, the energies of these two states are equal to each other,
as well to the energy of any state in which all variables $\varphi_{\bf
j}$ are equal either to $\Phi$ or to $\Phi+\pi$.
Therefore, in the absence of thermal fluctuations the
system would experience at $J_1=0$ a single phase transition
between the phases with different ordering.
Note that this property is not the consequence of keeping only two terms
in Eq. (\ref{V}) - for a more complex form of $V(\theta)$ the transition
will be shifted from the point where $J_1=0$ to the point where the two
minima of $V(\theta)$ have equal depths. However, it turns out that
in the presence of thermal fluctuations the single-transition scenario
does not survive.

{The finite temperature phase diagram} of the $XY$ model with a modified
{Berezinskii}-Villain interaction whose main features are
analogous to those of Eq. (\ref{V}) with $J_1,J_2>0$ has been constructed
in Refs. {}\cite{K85}.
In terms of the SFS array problem with interaction (\ref{V}) the main
conclusions of these works (confirmed in numerical simulations of Ref.
{}\cite{CCh}) can be reformulated and generalized as follows.

When both $J_1$ and $J_2$ are positive and much larger than $T$, the system
is in the phase with an algebraic decay of the correlation function
\begin{equation}                                            \label{C1}
    C_1({\bf j}_1-{{\bf j}_2})=
    \langle\exp i(\varphi_{{\bf j}_1}-\varphi_{{\bf j}_2})\rangle\,.
 \end{equation}
For brevity we shall call this phase ferromagnetic, although more
accurately it should be called a phase with algebraically decaying
ferromagnetic correlations. But since in two-dimensional systems with a
continuous order parameter the real long-range order is impossible
\cite{MW} and an algebraic decay of correlations \cite{W67} is as much as
one can get, the application of such a shorthand is rather natural. In
terms of SFS array this phase is superconducting and is characterized by a
finite superfluid density.

The decrease of $J_1$ down to $J_1\sim T$ induces a phase transition of
the Ising type related to the proliferation of solitons (a soliton is a
linear topological excitation on crossing which the phase jumps by $\pi$).
The existence of such a transition is especially evident for $J_2=\infty$
when the model defined by Eqs. (\ref{H}) and (\ref{V}) is reduced to the
Ising model with coupling constant $J_1$, however it exists (and has the
same nature) also when $J_2$ is less than infinite. The proliferation of
solitons leads to the replacement of the algebraic decay of the
correlation function $C_1({\bf r})$ by an exponential one. On the other
hand, on both sides of the transition the superfluid density remains
finite, which for $J_2<\infty$ manifests itself in the algebraic decay of
the correlation function
\begin{equation}
C_2({\bf
j}_1-{{\bf j}_2})=
    \langle\exp 2i(\varphi_{{\bf j}_1}-\varphi_{{\bf j}_2})\rangle\,.
\end{equation}

It is clear that in the phase with such a behavior of $C_1({\bf r})$ and
$C_2({\bf r})$ the role of the order parameter is played by $\exp
(2i\varphi_{\bf j})$ and therefore formally it can be called nematic.
Analogous nematic phase (induced by the proliferation of solitons) is
expected to exist in thin films of superfluid $^3$He \cite{K85b}. In the
nematic phase of a SFS array, the superconducting current can be
associated with the motion of pairs of Cooper pairs and therefore this
phase can be identified by studying the periodicity of the persistent
current in the array with annular geometry penetrated by a magnetic flux
(the period has to be equal to half of the superconducting flux quantum).

The relevant topological excitations in the nematic phase are
halfvortices, that is the vortices with topological charges $\pm 1/2$
which are the end points of solitons.
The interaction of these objects 
is logarithmic and keeps them bound in pairs, which allows one to treat
solitons as closed lines playing the role of domain walls in the Ising
model. With decrease in $J_2$ the strength of the logarithmic interaction
of half-vortices goes down and at $J_2\sim T$ it becomes too weak to keep
them bound in pairs. The phase transition related to the dissociation of
bound pairs of half-vortices is of the 
{Berezinskii}-Kosterlitz-Thouless (BKT) type. It differs from the standard
BKT transition by the value of the superfluid density jump, which is
larger by the factor of 4. In the disordered phase the superfluid density
vanishes and correlation function $C_2({\bf r})$ also decays
exponentially.

For $J_2 \ll T$ the disordered phase is separated from the ferromagnetic
phase existing at large enough ratio $J_1/T$ by the standard BKT
transition related with the dissociation of pairs of integer vortices
(exactly like at $J_2=0$). With the decrease in the ratio $T/J_2$ one
encounters a tricritical point, where this BKT transition is transformed
into a first-order one (with larger than universal jump of the superfluid
density). A change in the nature of the transition can be associated with
switching on of a different mechanism for the destruction of the
ferromagnetic ordering. On the other side of the tricritical point the
disordering is triggered not by the integer vortex pair unbinding but by
the proliferation of solitons taking place when the logarithmic
interaction of half-vortices is too weak to keep them bound in pairs. This
induces the simultaneous unbinding of integer vortices which takes place
not because their direct logarithmic interaction is insufficiently strong
but because it is screened by the presence of free half-vortices.

\begin{figure}
\onefigure{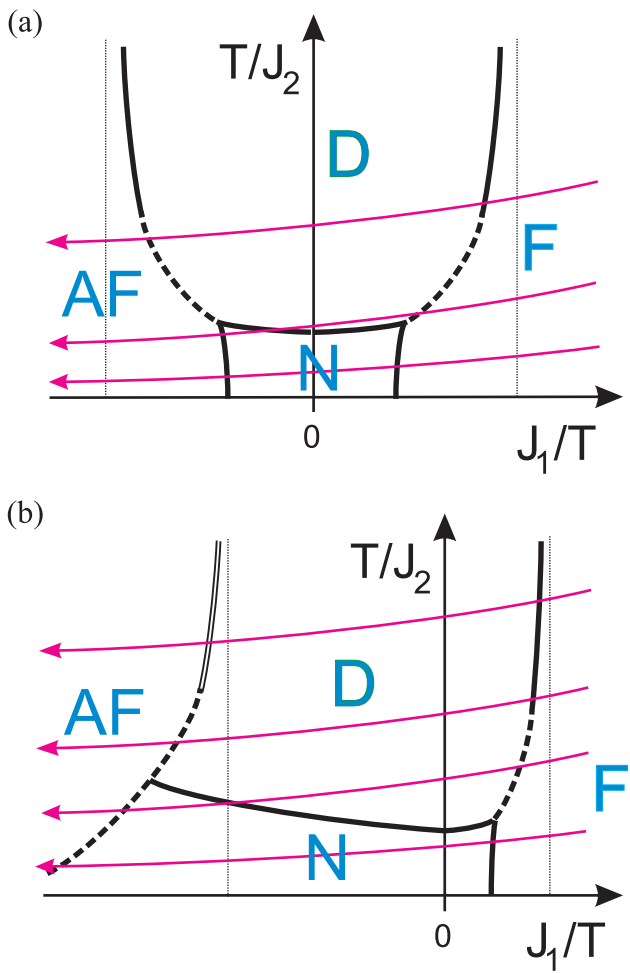}
\caption[Fig. 1]  {(Color on-line) Schematic structures of phase diagrams
of SFS arrays with (a) bipartite lattice and (b) triangular lattice.
Ferromagnetic (F), nematic (N), antiferromagnetic (AF) and disordered (D)
phases are separated from each other either by continuous (continuous bold
lines) or first-order (dashed bold lines) phase transitions. The double
line separating AF and D phases in (b) stands for the sequence of BKT and
Ising transitions with very small separation. Curved arrows going from
right to left show different paths of the evolution of an array with the
decrease in temperature. }\label{fig1}
\end{figure}
\vspace*{2mm}

The schematic structure of the phase diagram containing ferromagnetic (F),
nematic (N) and disordered (D) phases is shown 
{in Fig. \ref{fig1}(a) in coordinates $J_1/T$ and $T/J_2$. Although the
above analysis refers only to the right half of this figure (with
$J_1/T>0$), in the case of} a SFS array with a bipartite lattice it is
clear from the symmetry of the problem that at negative values of $J_1/T$
the phase diagram has exactly the same form as at positive, the only
difference being that the phase with the ferromagnetic algebraic
correlations is replaced by the phase with the antiferromagnetic algebraic
correlations (which have the two-sublattice structure).

The evolution of a SFS array with the decrease in temperature is shown in
Fig. \ref{fig1}(a) by curved arrows going from  right to left. From the
structure of the phase diagram it is clear that when thermal fluctuations
are taken into account the direct phase transition between the
ferromagnetic and antiferromagnetic phases is no longer possible and is
replaced by a finite region containing either one or two intermediate
phases.

In particular, for sufficiently low values of $T_0/J_2(T_0)$ the evolution
goes along the path \makebox{F-N-AF}, that is, the ferromagnetic and
antiferromagnetic phases are separated by the nematic phase, both phase
transitions being of the Ising type. On the other hand, for sufficiently
high values of $T_0/J_2(T_0)$ the ferromagnetic and antiferromagnetic
phases are separated by the strip of the disordered phase and the phase
transitions are either of the BKT type or of the first order. For
intermediate values of $T_0/J_2(T_0)$ the evolution has to take place
along the path \makebox{F-D-N-AF} involving three different phase
transitions and if in the region where $|J_1(T)|$ is comparable with $T$
or smaller the ratio $T/J_2(T)$ changes extremely little (by less than few
percent), the path \makebox{F-N-D-N-AF} involving {\em four} phase
transitions is also possible, although it hardly can be called a typical
one.

\section{Triangular lattice}
In the case of a {triangular} lattice the structure of the phase
diagram at $J_1>0$ is basically the same as for a bipartite
lattice, whereas at $J_1<0$ the situation is essentially different.
The main reason for that is that at negative $J_1$ the structure of the
ground state is different for small and for large values of $|J_1|$. In
particular, for \makebox{$-9J_2<J_1<0$} the minimum of energy
is achieved when on each triangular plaquette the phase difference on two
bonds is equal to $\pi$ and on the third one to zero. It is clear that in
any configuration satisfying this rule the variables $\varphi_{\bf j}$ can
acquire only two values which differ by $\pi$ (for example, $\Phi$ and
$\Phi+\pi$), from where it follows that in terms of the nematic order
parameter $\exp(2i\varphi_{\bf j})$ the system is perfectly ordered.

After introducing bimodal variables $\sigma_{\bf j}=\pm 1$ (below they are
called pseudospins) such that
\begin{equation}                                             \label{phi}
\exp(i\varphi_{\bf j})=\exp(i\Phi)\sigma_{\bf j}\,,
\end{equation}
one finds that the above-mentioned rule is satisfied as soon as each
triangular plaquette contains both positive and negative pseudospins. This
means that the set of the allowed configurations of pseudospins
$\sigma_{\bf j}$ coincides with the set of the ground states of the
antiferromagnetic Ising model with triangular lattice (the AFMITL model).
The number of such configurations grows exponentially with the size of the
system
\cite{Wan}. The exact solution of the AFMITL model \cite{Wan,Hout} at zero
temperature is characterized by an algebraic decay of the correlation
functions \cite{Steph}, in particular,
\makebox{$\langle \sigma_{{\bf j}_1}\sigma_{{\bf j}_2}\rangle \propto|{\bf
j}_1-{\bf j}_2|^{-1/2}\!$.}
These correlations have the three-sublattice antiferromagnetic structure,
that is are positive when the two pseudospins belong to the same
triangular sublattice and negative \makebox{otherwise} \cite{Steph}.
From the form of Eq. (\ref{phi}) it is then clear that at zero temperature
$C_1({\bf j}_1-{\bf j}_2)$ coincides with $\langle \sigma_{{\bf
j}_1}\sigma_{{\bf j}_2}\rangle$ and therefore has a three-sublattice
antiferromagnetic structure.

At $J_1<-9J_2$ the ground state of (\ref{H}) has exactly the same
structure as at $J_2=0$. 
In this state each of the three sublattices is ferromagnetically ordered
but the phases in the different sublattices are rotated with respect to
each other by $\pm 2\pi/3$ \cite{MS}. The full set of ground states is
characterized by a combined $U(1)\times Z_2$ degeneracy, where $U(1)$
corresponds to the simultaneous rotation of all phases and $Z_2$ can be
associated with antiferromagnetic ordering of chiralities of triangular
plaquettes. Thus in the absence of thermal fluctuations the phase diagram
of a SFS array with triangular lattice would incorporate three different
phases, the phases with ferromagnetic and antiferromagnetic ordering being
separated by a wide strip of the phase with perfect nematic ordering and
an algebraic decay of antiferromagnetic correlations.

At finite temperatures the perfect antiferromagnetic ordering existing at
$J_1<-9J_2$ is naturally replaced by an algebraic decay of $C_1({\bf r})$,
however a finite superfluid density and the genuine long-range order in
staggered chirality survive. It is known both from numerical simulations
\cite{LL98} and analytical considerations \cite{K02} that at $J_2=0$ the
disordering of the system with the increase in temperature takes place
through the sequence of two phase transitions which are situated
very close to each other.
The first of them is related to vortex pairs dissociation and is of
the BKT type, whereas the second is related with domain wall proliferation
and is of the Ising type. It follows from the analysis of the mutual
influence of the topological excitations of different types \cite{K02}
that the same scenario can be expected to hold also when $J_2>0$.

The properties of the nematic phase are influenced by a small finite
temperature more drastically than that of the antiferromagnetic phase. It
is known both from the exact solutions \cite{Wan,Hout} and from the
mapping onto a solid-on-solid (SOS) model \cite{BH} that at any finite
temperature the isotropic AFMITL model is in the disordered phase with a
finite correlation radius (which diverges when $T\rightarrow 0$). This
immediately allows one to conclude that at $T>0$ the nematic phase is
characterized by an exponential decay of $C_1({\bf r})$. On the other
hand, spin wave fluctuations lead to an algebraic decay of $C_2({\bf r})$.
These properties are in perfect agreement with those of the nematic phase
at $J_1>0$, which is no surprise since this is just the same phase.
Exactly like at $J_1>0$, at $J_1<0$ the nematic phase is characterized by
a finite superfluid density and its disordering takes place via BKT phase
transition related to the dissociation of halfvortex pairs. One more
example of an XY model in which the phase transition into a disordered
phase is related to the dissociation of halfvortex pairs is the frustrated
XY model with dice lattice and one-third of flux quantum per plaquette
\cite{K05}.

Since at $-9J_2<J_1<0$ the nematic phase is characterized by a finite
residual entropy $S_0\approx 0.323$ \cite{Wan}, the first-order transition
line separating it from the antiferromagnetic phase at finite temperatures
is shifted to larger values of $|J_1|$ (in particular, at low temperatures
it takes place at
$J_1\approx -9J_2-2S_0T$).
Together with what we already know about the disordering of the
antiferromagnetic and nematic phases this allows us to draw the schematic
phase diagram for the case of a triangular lattice shown in Fig.
\ref{fig1}(b).

Like in Fig. \ref{fig1}(a), curved arrows going from right to left show
the evolution of the system with the decrease in temperature. The four
arrows present in Fig. \ref{fig1}(b) correspond (starting from the lowest
one) to scenarios \makebox{F-N-AF}, \makebox{F-D-N-AF}, F-D-AF, and F-D-C-AF,
respectively. Here C denotes the phase with long-range order in chirality
and vanishing superfluid density which separates AF and D phases at
sufficiently high values of $T_0/J_2(T_0)$. Like for a bipartite lattice,
the four-transition scenario (involving the path \makebox{F-N-D-N-AF})
is also possible if the region
where $|J_1(T)|$ is comparable with $T$ or smaller is sufficiently narrow.

\section{Conclusion} In the present letter we have investigated what happens
with a phase-coherent array of SFS junctions when the decrease of
temperature leads to the crossover of the junctions to the $\pi$-state.
The corresponding phase diagrams have been constructed for the cases of a
bipartite lattice (square or honeycomb) and of a triangular lattice.
We have shown that the transition from the coherent phase existing well
above the crossover to the coherent phase existing well below the
crossover is never direct and these two phases are always separated by one
or more intermediate phase(s). Naturally, the same approach can be used to
construct the phase diagrams in the vicinity of the second crossover (from
the $\pi$-state back to the conventional state) if it does exist.
We hope that our results will stimulate more active experimental
investigations of SFS junction arrays.

The Hamiltonian (\ref{H}) can be also used for the description of a planar
magnet with both bilinear and biquadratic exchange in the situation when
the biquadratic exchange is ferromagnetic. For the case of the
antiferromagnetic biquadratic exchange ($J_2<0$) such a system with a
triangular lattice has been investigated by Park \etal \cite{PONH}. Its
phase diagram also includes a nematic phase, which however has a more
complex structure of correlations (a three-sublattice one) than the
nematic phase discussed in this work.

\vspace*{2mm}

The author is grateful to {Ya. V. Fominov} for useful discussions. This
work has been supported 
by the RF President Grant for Scientific Schools No. 5786.2008.2.


\end{document}